# Investigation of Dependence between Time-zero and Time-dependent Variability in High-κ NMOS Transistors

Mohammad Khaled Hassan, *Student Member, IEEE* and Kaushik Roy, *Fellow, IEEE*

*Abstract*— Bias Temperature Instability (BTI) is a major reliability concern in CMOS technology, especially with High dielectric constant (High-κ/HK) metal gate (MG) transistors. In addition, the time independent process induced variation has also increased because of the aggressive scaling down of devices. As a result, the faster devices at the lower threshold voltage distribution tail experience higher stress, leading to additional skewness in the BTI degradation. Since time dependent dielectric breakdown (TDDB) and stress-induced leakage current (SILC) in NMOS devices are correlated to BTI, it is necessary to investigate the effect of time zero variability on all these effects simultaneously. To that effect, we propose a simulation framework to model and analyze the impact of time-zero variability (in particular, random dopant fluctuations) on different aging effects. For small area devices (~1000 nm$^2$) in 30nm technology, we observe significant effect of Random Dopant Fluctuation (RDF) on BTI induced variability ($\sigma_{\Delta Vth}$). In addition, the circuit analysis reveals similar trend on the performance degradation. However, both TDDB and SILC show weak dependence on RDF. We conclude that the effect of RDF on $V_{th}$ degradation cannot be disregarded in scaled technology and needs to be considered for variation tolerant circuit design.

## I. INTRODUCTION

IN order to aggressively downscale devices and suppress the standby leakage current, HfO$_2$ based HKMG transistors were adopted recently by the IC industry [1]. Their advantages over conventional silicon oxynitride (SiON) devices have already been demonstrated to a great extent in different literature [1-3]. However, this changeover in technology roadmap brings about new failure mechanisms (such as Positive Bias Temperature Instability (PBTI), Stress Induced Leakage Currents (SILC), etc.) that enhanced the complexity in reliability quantification [4-6]. Due to the presence of d-shell electrons and relatively higher coordination number, high-κ oxides are vulnerable to both native and time dependent stress induced defect formation [7-8]. Since HfO$_2$ bulk traps cannot be charged under negative stress voltage, these defects only affect the threshold voltage and leakage current of NMOSFETs [4]. Along with the time dependent BTI induced variation, the on -die/die-to-die and parameter variations play a critical role in reliability assessment at both device and circuit level [9-10]. In sub-45nm technology, the native oxide thickness has gone down to 2 or 3 atomic layers. Consequently, it is important to understand the nature of $\sigma_{\Delta Vth}$ with the change in device dimensions. In addition, the dependence between time-zero and time dependent variability needs to be properly addressed. The devices at the lower process induced distribution tail performs faster (Fig. 1) due to their low turn-on voltage compared to the nominal threshold voltage, $V_{th}(0)$. The variation can be up to few sigma values ($\sigma_{Vth(0)}$) of the time-zero variation. This increases the stress voltage across the high-κ and interfacial layer (IL), causing faster trap generation in the corresponding oxide layers. As an example, in order to comprehend the nature of RDF and temporal BTI evolution and their dependence, it is important to capture the effect of each dopant in the channel region and defect in the bulk oxide layers simultaneously and determine the corresponding effect on $V_{th}$ degradation. Therefore, a complete modeling framework for RDF as well as BTI variation considering the fluctuation in number and position of independent dopants and their impact on oxide defects/traps is necessary for device and circuit analysis.

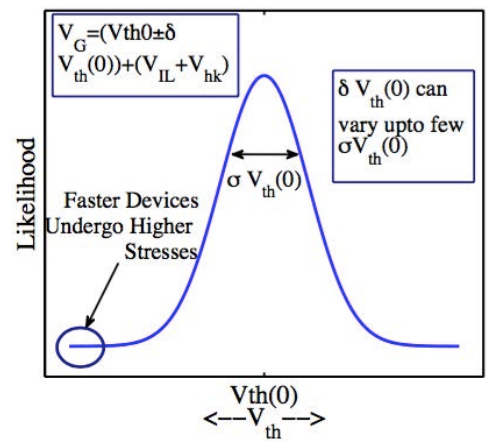

Fig. 1: Time-zero variation ($\delta_{Vth(0)}$) with respect to the nominal threshold voltage, $V_{th}0$) can be up to few $\sigma_{Vth(0)}$. This can cause a non-uniform stress voltage distribution across the oxide layers.

This work is funded in part by the National Science Foundation.

The authors are with the School of Electrical and Computer Engineering at Purdue University, West Lafayette, IN 47907 (email: khaled@purdue.edu, kaushik@purdue.edu)

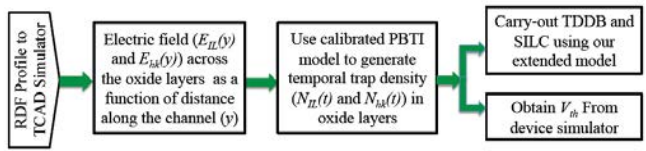

Fig. 2: Our simulation framework.

The rest of the paper is organized as follows. In section II, we explained our modeling and simulation framework. In this section, we have briefly introduced the PBTI model proposed in [11] and our extended model in order to analyze time dependent oxide wear outs. Section III discuses the possible dependence between RDF and BTI at the device and circuit level of design. We have also carried out TDDB and SILC analysis for NMOS devices considering this interrelation. Finally section IV concludes this work.

## II. FRAMEWORK FOR VARIABILITY ANALYSIS

Fig. 2 shows the framework of our analysis. We have investigated the dependence between time-zero variation (RDF) and time dependent variability and evaluated its impact on aging effects such as, BTI, TDDB, and SILC. Understanding and modeling such interrelation can be important in reliability aware circuit design. In order to properly model the impact of RDF on different aging effects, the non-uniformity of electric field across the dual oxide layers (HK and IL) along the channel length needs to be considered. To that effect, we have used a TCAD simulator to carry out our device level simulation [12]. We divided the channel region in $35x30$ number points and extracted the oxide electric field as a function of the channel length at each of the points along the width. The extracted fields are then incorporated to our BTI trap generation model [11] and calibrated against [13]. Subsequently, we integrated our PBTI model to the device simulator in order to analyze the dependence between RDF and PBTI. The TDDB and SILC analysis are carried using our extended trap generation model developed in MATLAB [14].

The NMOS device designed in the TCAD simulator is benchmarked with ITRS 2009 at the 30nm technology node [5]. The thicknesses of the high-κ and interfacial layers are 2nm and 0.6nm, respectively. Other design parameters are

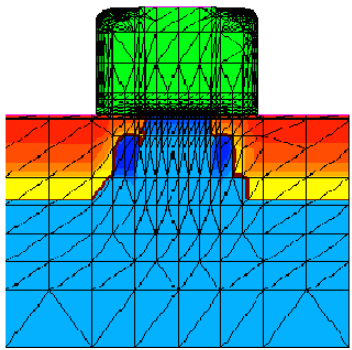

Fig. 3: The cross-section of our simulated NMOS device.

### Table 1: NMOS and PMOS design parameters

| Parameter Name | Value |
|---|---|
| Gate Length (L) | 30nm |
| Gate Width (W) | 35nm |
| Gate Oxide Material | $HfO_2$, $SiO_2$ |
| Equivalent Oxide Thickness | 0.95nm |
| Channel Doping | $2.7x10^{18}/cm^3$ |
| Source/Drain Doping | $6x10^{20}/cm^3$ |
| Halo Doping | $2x10^{19}/cm^3$ |
| Source/Drain Extension | 10nm |
| Spacer Length | 20nm |
| NMOSFET Gate Workfunction | 4.6eV |
| PMOSFET Gate Workfunction | 5.06eV |

shown in Table 1. The ON and OFF currents of the simulated NMOS device are 1190 μA/μm and 100nA/μm, respectively. We have used the drift-diffusion transport model and the unified mobility model from Philips for the simulations [15]. For quantum correction, the device simulator uses Van Dort model [16]. Detailed explanation of these models and a list of all the parameters used are available in the device user guide of the TCAD simulator [12]. Fig. 3 shows the cross-section of our NMOS device. Unless otherwise specified, the stress voltage and temperature were kept at 2V and 125°C, respectively. We applied a drain voltage of 50mV in the device simulator and extracted $V_{th}$ using the maximum-transconductance method, which is a built-in feature of the TCAD simulator [12]. Different parameters used in our reliability analysis are shown in Table.2. Below we discuss the modeling and analysis of RDF and different time dependent aging effects that we integrated in our simulation framework.

### A. Time-zero Variability

Time-zero fluctuation is comprised of three major components (1) Random Dopant Fluctuation (RDF), (2) Mean Gate Length Deviation (GLD), and (3) Line Edge Roughness (LER) [18]. Among these three, RDF is considered to have the most adverse effect [18-20]. With the aggressive scaling down of

### Table 2: Different parameters used in our simulation

| Parameters | Value | Reference |
|---|---|---|
| $\alpha(HfO_2)$ | 0.17 | [4, 13] |
| $\alpha(SiO_2)$ | 0.38 | [39] |
| $E_r(HfO_2)$ | 0.8eV | [40] |
| $E_r(SiO_2)$ | 0.1eV | [41] |
| $a_0(HfO_2)$ | 1nm | [17] |
| $a_0(SiO_2)$ | 0.6nm | [17] |
| $\Phi_B$ | 3.2eV | [31] |
| $\sigma_n$ | $10^{-14}-10^{-15}$ cm² | [27] |
| $a_0$=trap size $\sigma_n$: Trap Cross section $E_r$: Trap Relaxation Energy $\alpha$: Power exponent of time dependence $\Phi_B$: Barrier height seen by the electrons | | |

MOSFETs, the intrinsic variation of $\sigma_{Vth(0)}$ due to the smaller number of discrete implanted dopants and their corresponding random position in the channel becomes more distinct [20-21]. RDF causes a large variation among similar transistors and affects performance at the device as well as the circuit level of design abstraction [22]. In this work, we have generated RDF profiles considering randomness in the number of dopants using a pseudo-random number generator [14].

$$N_{RDF}^i(0) = Poiss(N_{RDF_{avg}}(0))_{1xS}, \text{ where } i = 1,2,....,S \quad (1)$$

where, $N_{RDF}^i$ is a randomly drawn number from an 1xS vector generated by Poisson's distribution. $N_{RDFavg}$ is the average doping density in the channel region and $S$ is the number of microscopically similar sample devices. In addition to the number of random dopants, randomness of the location of the dopants also plays a significant role on $V_{th}$ variation. Dopants that are closer to the Si/SiO$_2$ interface have larger influence on the $V_{th}$ fluctuation [21]. Moreover, Drain Induced Barrier Lowering (DIBL) increases the standard deviation of $V_{th}$ in nanoscale transistors that is attributed to RDF. This is because the drain potential can nullify the effect of dopants that are located close to the drain contact and a large number of dopants close to the drain contact can cause high potential profile [22]. The RDF profiles are fed to the TCAD simulator and the locations of the dopants were randomized using the built-in feature of the TCAD simulator.

### B. Positive Bias Temperature Instability (PBTI)

PBTI emerged as a significant reliability concern for High-κ NMOS transistors. We proposed a physics based statistical PBTI model in [11] where the location of each of the traps were considered separately. The trap generation process in dielectric layers depends on stress time (t), temperature (T), and, the applied gate bias (V$_G$) and the number of generated traps ($N_G$) can be modeled as follows:

$$N_G(V_G, T, t) = A \times (V_G)^\gamma \exp(-E_a / K_B T) t^n \quad (2)$$

where, $V_G$ is the voltage drop across the dielectric layer, $\gamma$ is the voltage acceleration factor, $n$ is the PBTI power exponent, $K_B$ is the Boltzmann constant, and $A$ is a proportionality constant and the calibration parameter that depends on the device structure. Typical values of n and $\gamma$ for PBTI degradation are ~0.17 and ~5, respectively [4]. We assumed a random distribution of these traps across the dielectric layers and assigned a stochastic process $S(t)$ to each of the oxide bonds:

$$S(t) = \begin{cases} 1: \text{if the bond is broken} \\ 0: \text{if the bond is tied} \end{cases} \quad (3)$$

Therefore, the temporal variation of randomly distributed generated traps can be given as

$$N_{(Gi)_t}(t) = \int_{x=0}^{W} \int_{y=0}^{L} \int_{z=0}^{T_d} S(t) dx dy dz \quad (4)$$

where, $W$, $L$, and $T_d$ are respectively the width, length, and, the thickness of the oxide layer. For our simulation, we have discretized the oxide layers along their width, length, and height depending on the defect sizes in the corresponding oxide layers (Table 2).

Using probability theory, we have determined the number of charged traps in the oxide layers. Trapping of carriers is dependent on the generation of traps and therefore is a conditional event. The probability that both trapping and trap generation events can occur at the same location ($x$, $y$, $z$) at a specified time is given by [23]:

$$P(C | G) = \frac{P(C \bullet G)}{P(G)} \quad (5)$$

where $P_G$ and $P_C$ represent the probability of trap generation and trap charging respectively, $G$ and $C$ are trap generation and charging events respectively, and $P(C \bullet G)$ is the probability of $C$ and $G$ taking place at the same time. $P(C | G)$ in (5) is the probability of charging a trap at a location (x,y,z) given a trap is being generated at the same location. Since $C$ is a subset of $G$, (5) can be simplified to

$$P(C | G) = \frac{P(C)}{P(G)} \quad (6)$$

In order to model the conditional probability $P(C|G)$, we calculated the tunneling probability of electrons, $P_{WKB}$ by *Wentzel–Kramers–Brillouin* (*WKB*) method [24] and compared that with a set of pseudo-random probabilities $P_r$.

$$P_{WKB} = \exp\left(-\frac{2}{\hbar}\left[\int_0^z \sqrt{2m_{de}^*(\phi_{Be} - E_k)} \, dz\right]\right) \quad (7)$$

where, $m_{de}^*$, $\phi_{Be}$, $\hbar$, $E_k$ are the effective mass of electron in dielectric layers, effective barrier height seen by the tunneling electron, reduced Planck's constant, and the average kinetic energy of electron in a 3D semiconductor, respectively. $P_r$ is generated in such a way that they have the same order of magnitude as that of $P_{WKB}$. Once we have $P_r$ and $P_{WKB}$, $P(C|G)$ can be modeled as follows:

$$P(C | G) = \begin{cases} 1: \text{if } P_{WKB} > P_r \\ 0: \text{Otherwise} \end{cases} \quad (8)$$

Since we can extract the locations of the generated traps from our simulation, $P(G)$ is either $0$ or $1$ if a bond is tied or broken, respectively. In addition, $P(C)=0$ if the bond is tied. Therefore, at a certain time $t$, (6) becomes:

$$P(C)_t = P(C | G)_t \quad (9)$$

We calculate the total number of charged traps ($N_c$) using the following expression:

$$N_C(t) = \int_{x=1}^{W} \int_{y=1}^{L} \int_{z=1}^{T_d} P(C)_{t,x,y,z} \, dx \, dy \, dz \quad (10)$$

From the total number of charged traps $N_C(t)$, $\Delta V_{th}$ can be determined using poison's equation [11] or with the aid of the TCAD simulator [12].

The total $V_{th}$ degradation due to PBTI is dominated by the trap generation in HK layer since trap generation rate in IL layer is very small and becomes significant only at a very high stress condition. We have assumed a Poisson distributed number of traps among a large number of samples in this work. For detailed analysis, the readers are referred to [11].

### C. Time Dependent Dielectric Breakdown (TDDB)

The time dependent dielectric oxide breakdown (TDDB) in HKMG transistors is characterized by short breakdown times as well as shallow Weibull slopes [17]. We have incorporated a 3D percolation model in our BTI model in order to capture the TDDB effect considering different defect generation rates in the IL and the HK layers. Although, the high-κ materials demonstrate lower tunneling current and better transistor performance, a thin interfacial layer (IL) of $SiO_2$ is needed to passivate the Si-dielectric interface [17]. Since the defect generation rate in $SiO_2$ at operating voltage is lesser than the high-κ layer [13], the presence of this IL layer improves both TDDB and PBTI and also increases the carrier mobility [13, 17].

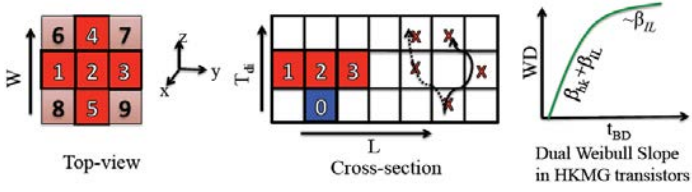

Fig. 4: Implementation of the 3D percolation model. One single defect in the bottom layer can lead to multiple breakdown paths. 'X' indicates a defect generated in a cell. Dual Weibull slope arises due to the different defect generation rates in the HK and IL layers.

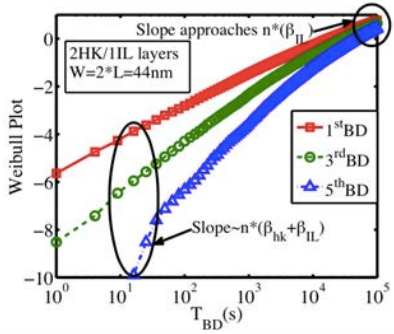

Fig. 5: Dual Weibull slope is observed because of the inhomogeneous defect generation rates in the HK and IL layers [17].

The formation of conduction paths over time due to the accumulation of oxide traps triggers the breakdown of a dielectric layer. The widely accepted percolation model in [25] can capture this idea for thick oxide layers. However, in scaled technologies, the formation of 3D conduction paths can no longer be ignored since the oxide layers are comprised of very few layers of defects. Subsequently, 3D cell-based percolation model has been proposed and implemented in different literature [17,26]. In our statistical model, we have incorporated this model in-order to capture the TDDB effect. The Weibull plot is determined as:

$$W = \ln(-\ln(1-F)) \qquad (11)$$

where, $F$ is the fraction of failed devices in a Monte Carlo (MC) event. Fig. 4 shows the cross section and top-view of a three-layer dielectric material and the dual Weibull slope observed in HK gate stacks. We have considered all the nearest neighbors in determining a possible link between two adjacent layers. In Fig. 4, a defect '0' in the bottom layer can have nine possible connections to the layer above itself. Defects '1', '2', '3', '4', and '5' have at least one common edge with '0' while defects '6', '7', '8', and '9' have exactly one node in common with '0'. When a continuous path is established between the top and the bottom layers, a conduction or breakdown path is considered to have formed. In case of dual-layer stack that is widely used in today's technology, the TDDB analysis has become more complicated since the high-κ and the interfacial layers have inhomogeneous defect generation rates. High-κ layers are prone to the formation of both native and stress induced defects. Hence, they wear out relatively faster than the IL layer and the ultra-thin IL layer determines the final catastrophic breakdown [27].

In order to validate the model, we applied a very high stress condition and ran our MC simulation with 15000 samples for $10^5$ seconds. We observe from Fig. 5 that the slopes for different breakdowns, although initially depend on the defect formation in both HK and IL layers, gradually become independent of the HK layer as the stress time increases. The Weibull slopes $\beta$ in the dielectric layers is defined as [17]

$$\beta = \alpha(\frac{T_{ox}}{a_0}) \qquad (12)$$

where, $\alpha$ is the time exponent for defect generation, $a_0$ is the cell or defect size, and $T_{ox}$ is the thickness of either HK or IL layer. We observe in Fig. 5 that the Weibull plot slowly approaches the slope in IL layer since the final hard breakdown of the devices is dictated by the trap generation in the IL layer. However, the actual reason behind the bi-modal distribution observed by different groups is still controversial. Yew, et al [28] experimentally demonstrated that the presence of grain boundary (GB) defects in the HK layer can cause early breakdowns in many devices and can give rise to similar bi-modal Weibull distribution. We have also considered this effect in our simulation later in results and discussion section.

### D.  Stress-Induced Leakage Current (SILC)

SILC is a means to analyze the buildup of oxide traps under BTI stress. As the trap density reaches a critical value, the SILC current can cause significant power consumption leading to hard breakdown of a device [29]. Therefore, it is necessary to understand the evolution of SILC over time and have appropriate model that can predict progressive wear outs of the oxide layers in a scaled technology. Fig. 6 shows the band diagram of an NMOS under inversion mode and the components of gate leakage currents. There are three major components of gate leakage [30]:

1.  Direct Tunneling Current ($J_{DT}$) or the time-zero gate leakage
2.  Trap-Assisted-Tunneling Current in preexisting traps ($J(TP)$)
3.  Stress Induced Leakage Current ($J_{SILC}(TG)$) through Trap-Assisted-Tunneling in generated traps

The time-zero leakage current is known as direct tunneling current ($J_{DT}$) while the stress-induced leakage current ($J_{SILC}$) is known as Trap-Assisted-Tunneling (TAT) current. $J_{SILC}$ starts increasing when a device is under electrical stress. We implemented the modified direct tunneling current for high-κ gate stack proposed in [31].

$$J_{DT} = \frac{q^3}{8\pi h \phi_B} \frac{\phi_B}{V_{SiO_2}} (\frac{2\phi_B}{V_{SiO_2}} - 1) \frac{V_{SiO_2}}{t_{SiO_2}} T_{WKB} \qquad (13)$$

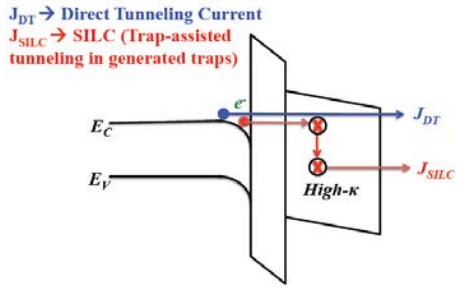

Fig. 6: Trap assisted SILC in HK layer decreases due to the high relaxation energy.

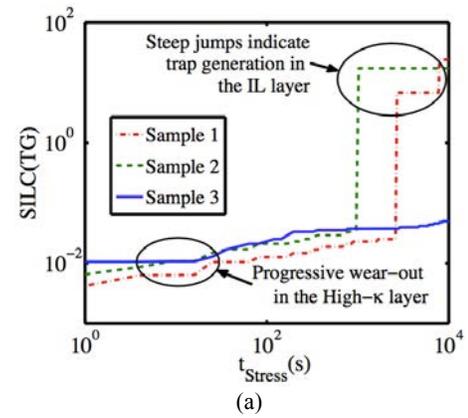

(a)

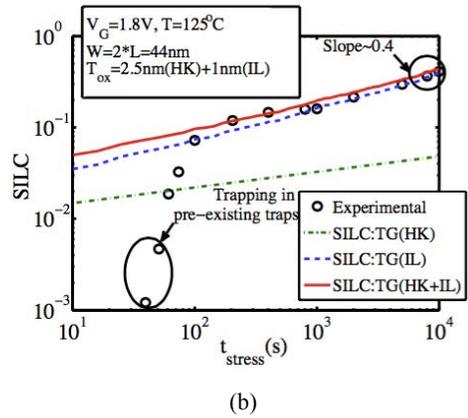

(b)

Fig. 7: (a) SILC in different small area sample devices indicates that trap generation in IL layer dominates total SILC (b) The average SILC in 15000 such devices considering only trap generation is in close agreement with the experimental data [13].

where, $\Phi_B$ is the barrier height seen by the carriers in conduction band of the substrate and $V_{SiO2}$, $t_{SiO2}$, and $T_{WKB}$ are the voltage drop across the SiO$_2$ layer, thickness of the IL layer, and the tunneling probability across the dual gate stack determined using WKB approximation [24], respectively. The stress-induced current due to trap generation (TG) is modeled as [13]

$$J_{TG}(t) = \iint k\sigma_n V_{th} N(t) T_r (f_c - f_a) dx dE \qquad (14)$$

$$T_r = \frac{T_{in}(E) T_{out}(E - E_r)}{T_{in}(E) + T_{out}(E - E_r)}$$

where, $T_{in}$ and $T_{out}$ are the tunneling probabilities to and from the traps, respectively, $E_r$ is the trap relaxation energy, $f_c$ and $f_a$ are the occupation factors of cathode and anode, respectively, $\sigma_n$ is the electron capture cross section of the traps, and $v_{th}$ is the thermal velocity. $k$ in (6) depends on electron concentration of conduction band and energy suppression factor. In our work, we have calibrated $k$ in order to fit with

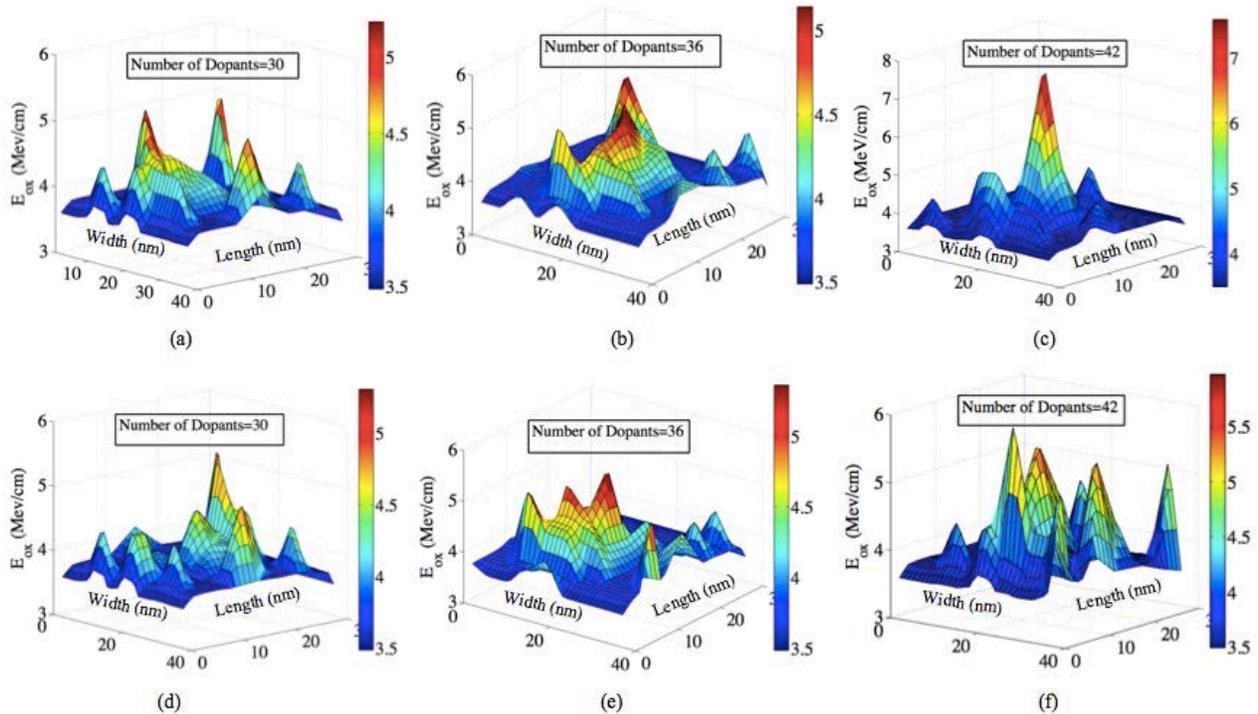

Fig. 8: Effect of random dopant fluctuation on the electric field across the IL oxide layer: (a), (b), and (c) show the electric field variation due to the fluctuation of the number of dopant in the channel region of an NMOS device. In comparison to the upper three figures, (d), (e), and (f) show the variation due to the fluctuation in dopant locations. A gate voltage of 1 V is applied for these simulations.

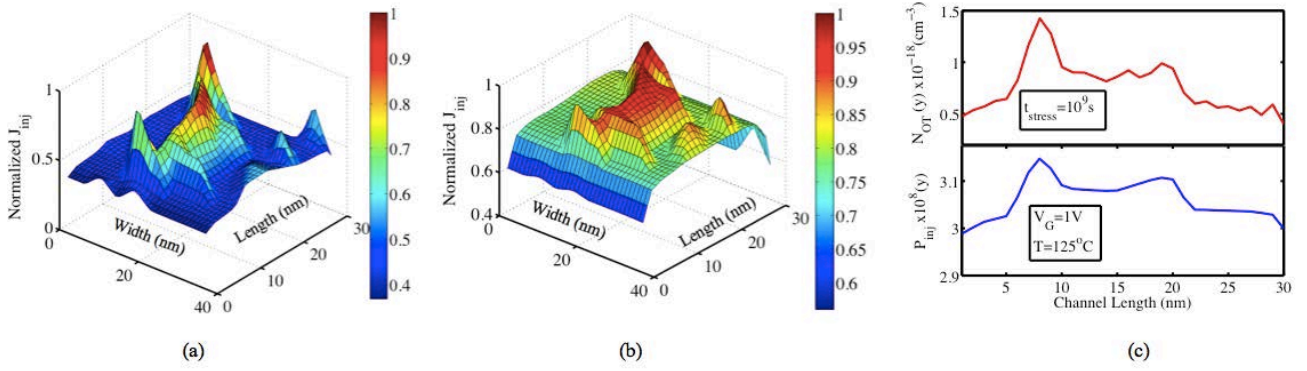

Fig. 9: Distribution of local injection current considering 36 dopants inside the channel region (a)$V_G$=1V, (b) $V_G$=2V (c) Correlation between local injection probability ($P_{inj}$) and PBTI trap density ($N_{OT}$) along the channel at $V_G$=1V. The results indicate direct correlation between RDF and PBTI. The correlation is expected to be higher at lower gate voltage.

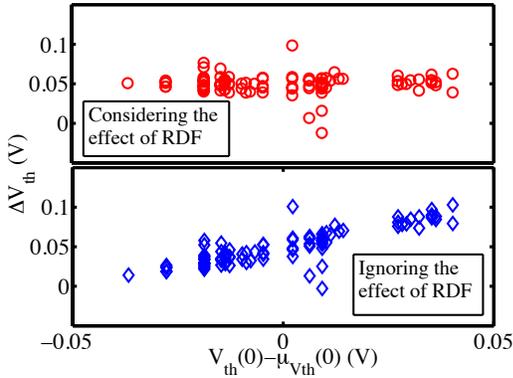

Fig. 10: Effect of RDF on PBTI at a mean $\Delta V_{th}$ of 50mV. Ignoring the effect of time zero variability leads to erroneous results in BTI estimation. The simulation was carried out at 2V and 125°C.

the experimental data in [13]. The normalized SILC current is then defined as [32]

$$J_{SILC}(t) = \frac{J_{TG}(t) - J_{TP}(t_{sat})}{J_0} \quad (15)$$

Here $t_{sat}$ is the saturation time of pre-existing trap filling which is usually in sub-milliseconds [33] and $J_0$ is the gate leakage current at $t$=0 which includes the direct tunneling component and tarp assisted tunneling in the pre-existing defects. The actual reason behind the SILC is a controversial issue. While some groups attribute the SILC phenomenon to the defect generation in the HK layers [30, 34, 35], others explain SILC by defect generation only in the IL layer [27, 36]. J. Yang, *et al* [13] argued that SILC is caused by defects in both HK and IL layers but mostly dominated by the defects in the IL layer since they have much lower relaxation energy. In our model we have considered degradation in both HK and IL layers and calibrated our model against [13]. Fig. 7 shows the average SILC current for a MC simulation of 15000 similar devices. As we can observe from Fig. 7a, for different small area sample devices, total SILC due to trap generation jumps up when there is a defect generation in the IL layer. This is because the relaxation energy in an IL defect is much lower than that of an HK defect [13]. The higher relaxation energy in the HK layer defects can exponentially decrease the tunneling out ($T_{out}$) probability and hence, can significantly decrease the contribution to the overall SILC. As a result, the slope of the average SILC (Fig. 7b) gradually approaches the time exponent of trap generation in IL layer. We have ignored the contribution from pre-existing traps and assumed $J_{TP}$=0 for our analysis since the trap density is not provided in [13]. Besides, these technology dependent native traps can be optimized to a very low degree [37] in which case only the trap generation plays a role in the SILC measurements.

### III. RESULTS AND DISCUSSIONS

In this section, we explain the impact of RDF on time dependent degradation using the results of our simulation framework. We extracted the electric field from the TCAD simulator. Fig. 8 shows the electric field variation across the IL layer among different devices with an average doping density of $2.7 \times 10^{18}$/cm³ in the channel region at $V_G$=1V. We see variations in the oxide electric field due to the randomization of dopant locations as well as the numbers. As a result of this, we expect to see the impact of RDF in stress induced trap density profiles among a large number of microscopically similar devices. We incorporated the extracted electric field in our PBTI model [11] in order to further investigate the impact of RDF on different aging effects. In Fig. 9, we have plotted the local gate injection current as a function of the channel length and width. The local injection current is determined using the direct tunneling current model in equation (13) and we plotted the normalized value (normalized with respect to the maximum value) for gate voltage of 1 and 2V in Fig. 9(a) and 9(b), respectively. For this simulation, we have considered the dopant profile corresponding to Fig. 8(a) where 36 dopants are randomly assigned inside the channel. We can see that RDF causes variation in the local injected current both at 1V and 2V. In order to investigate the relative impact of RDF, we determined the standard deviation of the local injected currents as a percentage of their average values. We observe that the variability induced by RDF at 1V (21.9%) is much higher compared to that at 2V (10.3%). Therefore, the impact is expected to be more significant at the device operating voltage. Since, PBTI is driven by the gate injection current, we investigated its correlation to local injected current probability ($P_{inj}$) in Fig. 9(c). For $V_G$=1V (corresponding to Fig. 9a), We extracted the probability as a function of the channel length at a random location along the width and incorporated the corresponding electric field distribution in our PBTI model. We simulated a large area device (Area =10000x30 nm$^2$) for a stress period of $10^9$ s and plotted the lateral distribution of the generated trap density in the upper panel of Fig 9(c). In the lower panel, we have plotted the corresponding local injection probability and observe that they are closely correlated. Therefore, we conclude from Fig.9 that the impact of RDF can modulate the $\Delta V_{TH}$ distribution induced by PBTI.

Fig. 10 demonstrates the impact of RDF on PBTI estimation for 50mV of mean $\Delta V_{th}$ degradation. The lower panel represents the data where the BTI degradation is estimated using an average time-zero threshold voltage, whereas the upper panel shows the same data taking the effect of RDF into account. The $\Delta V_{th}$ after a stress period of $t_{stress}$ for these two panels are defined as:

$$\Delta Vth(t_{stress}) = \begin{cases} Vth(t_{stress}) - Vth(0), & upper\ pannel \\ Vth(t_{stress}) - \mu_{Vth(0)}, & lower\ pannel \end{cases} \quad (16)$$

Both sets of data are plotted against the difference between the time-zero threshold voltage, $V_{th}(0)$ for each of the devices and the mean of time zero threshold voltage, $\mu_{Vth}(0)$. In this plot, the devices on the left side are relatively faster and we see that $\Delta V_{th}$ for these devices are higher in the upper panel. The effect of RDF on the devices that have threshold voltages close to the mean $\mu_{Vth}(0)$ are almost negligible. The devices on the right side of the panels on the other hand have higher time-zero threshold voltage. We see that the $\Delta V_{th}$ in the upper panel is relatively less than that of the lower panel. In other words, RDF causes higher degradation to the faster devices and lesser degradation to the slower devices as we have explained in Fig. 1.

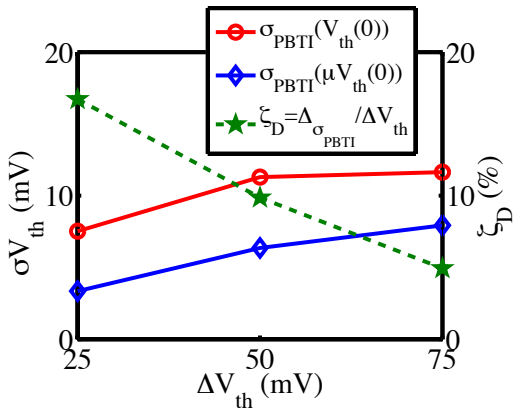

Fig. 11: Effect of RDF on PBTI decreases with the increase of BTI degradation.

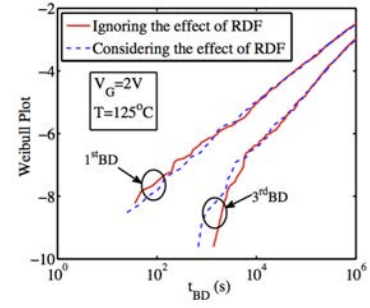

Fig. 12: The effect of RDF has very negligible or no impact on the TDDB estimation.

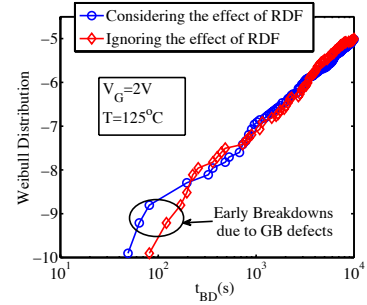

Fig. 13: Impact of RDF on TDDB taking the GB defects into account. The effect of RDF on TDDB is still negligible.

Fig. 11 shows the impact of RDF on the threshold voltage degradation at 25mV, 50mV, and 75mV of PBTI degradation. $\sigma PBTI(V_{th}(0))$ is the PBTI standard deviation taking the effect of RDF into account while $\sigma PBTI(\mu V_{th}(0))$ represents the PBTI variability considering average time zero $V_{th}$. We defined a dependence parameter $\zeta_D$, which is the underestimation of $\sigma_{\Delta Vth(PBTI)}$ with respect to mean $\Delta V_{th(PBTI)}$ when the effect of RDF is ignored and plotted on the right y-axis of Fig. 11. We observe that the RDF effect on the threshold voltage degradation decreases with the increase of BTI effect. This trend is in agreement with that of the experimental data presented in [9]. This observation can be significant for scaled technology where the margin for parametric variation is very stringent [38].

Next, we investigated the effect of RDF on TDDB and SILC at a stress voltage of 2V. The analysis was carried out for 15000 sample devices. Fig. 12 shows the Weibit plot for both the 1$^{st}$ and the 3$^{rd}$ breakdown with and without considering the effect of RDF. We observe negligible difference in TDDB lifetime. We carried out similar analysis taking the presence of GB defects into consideration. We divided the HK oxide layer in 42 grain-regions and assigned an average pre-existing GB defect density of 10$^{19}$/cm$^3$ in 25% of the sample devices. Since the presence of GB defects compromises the underlying IL layer [28], we assumed 10 times higher IL layer defect generation rate in these devices. The simulation results for 3$^{rd}$ soft breakdown is plotted in Fig. 13 and we observe negligible effect of RDF on TDDB. Early breakdowns due to GB defects in some of the devices gives rise to a bi-modal Weibull distribution, which is consistent with the observations reported in [28]. Since TDDB requires higher stress condition and defect density than BTI, the effect of RDF is negligible in both Fig. 12 and 13.

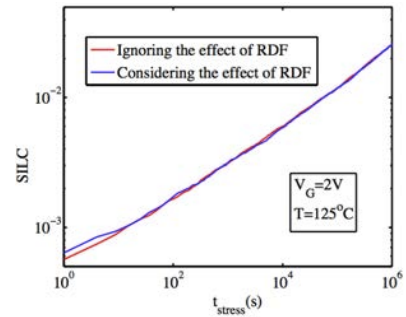

Fig. 14: Average SILC shows similar trend to TDDB since both of them are dominated by the trap generation rate in the IL layer.

Table 3: NBTI model used in our work

| |
|---|
| The average number of interface traps generated is predicted by the following H-$H_2$ R-D model [42]<br><br>$N_{IT_{avg}}(t) = WL(k_H/k_{H_2})^{1/3}(k_f(y)N_0/k_r)^{2/3}(6DH_2 t)^n$  (16)<br><br>$k_f(y) = k_{f0}Ec\exp(\gamma Eox(y))\exp(-\frac{EA_{k_f}}{qV_T})$   (17)<br><br>Where, $k_f$, $k_r$, $N_0$, $k_H$, $k_{H2}$, and $D_{H2}$ are Si-H bond breaking rate, Si-H bond annealing rate, the pre-stress Si-H bond density at the Si/SiO$_2$ interface, the generation rate, dissociation rate, and the diffusion coefficient for H and H$_2$. y in the above equation is the distance along the channel length. $k_{f0}$ is the calibration parameter. |
| In the above equation,<br><br>$$X = X_0 e^{-\frac{EA_X}{qV_T}}$$    (18)<br>$$X = \{k_f, k_r, k_H, k_{H_2}, D_{H_2}\}$$<br><br>where, $EA_X$ are the activation energies and $X_0\{=k_{f0}, k_{r0}, k_{H0}, k_{H20}, DH_{20}\}$are constants |
| n=1/6, $\gamma$= 0.6 ± 0.05, EF=0.36eV [42]<br>EA$_{kf}$=0.175eV,  EA$_{kr}$=0.2eV,  EA$_{kH}$=EA$_{kH2}$=0.3eV,<br>EA$_{DH2}$=0.58eV,  K$_{r0}$=9.9x10$^{-7}$,  D$_{H0}$=9.56x10$^{-11}$cm$^2$/s,<br>K$_{H0}$=8.56cm$^3$/s, K$_{H20}$=5.7x10$^5$/s, D$_{H20}$=3.5x10$^{-5}$cm$^2$/s [45] |

A similar trend is observed for SILC analysis and is shown in Fig. 14. We observe almost indistinguishable SILC for both average doping effect and RDF. The reason being, SILC is dominated by the trap generation in the IL layer where the defect generation rate is assumed to be very small in our simulations. Due to the low density of generated traps in the IL layer, the statistical fluctuation due to the RDF effect has trivial effect on SILC. SILC is highly a controversial issue and the relative contribution of stress-induced traps in the HK and IL layers is still debatable. Therefore, further investigation on the effect of RDF on SILC is necessary.

Finally, we investigated the effect of RDF on BTI degradation at the circuit level and analyzed the impact of correlation between RDF and BTI on the degradation in frequency of oscillation ($f_o$) and transient supply current ($I_{DDT}$) of ring oscillators. We extracted I-V characteristics from the TCAD simulator and implemented a look-up table based Verilog-A model for both n and p MOSFETs. Using the models, we carried out HSPICE Monte Carlo simulations on 10000 samples. For NBTI degradation in PMOS transistors, we implemented the reaction-diffusion (RD) model [42] and assumed Poisson's distribution on the interface trap numbers among different microscopic samples [43]. The average number of interface traps was calibrated against [44]. The RD model used in our work is summarized in Table 3. We have used the same framework as shown in Fig. 2 to take the effect of RDF on NBTI into account. In order to explore the performance degradation of the relatively slower circuits, we determined the -3$\sigma$ points of the above-mentioned parameters with and without considering the effect of RDF on BTI and plotted their difference as a percentage of the standard deviation ($\sigma$) in Fig. 15. We observe that ignoring the effect of RDF causes significant underestimation of both $f_o$ and $I_{DDT}$ especially at a relatively lower BTI degradation. Similar to the device level results, the influence of RDF decreases with the increase in BTI degradation. Therefore, the effect of RDF on BTI may have to be considered in scaled technology where the variability margin is strict.

## IV. CONCLUSION

In this paper, we have proposed a statistical modeling framework that takes the dependence between time-zero variability and different time dependent aging effects into account. The framework is developed based on our proposed work on PBTI, a statistical RD model on NBTI, a 3D cell based percolation model on TDDB, and a TAT current model on SILC. We have analyzed small area devices (~1000nm$^2$) and shown that 1) RDF has considerable impact on BTI effect and 2) the influence of RDF on time dependent oxide breakdown is almost negligible. In addition, we carried out circuit analysis using a look up table based Verilog-A SPICE model and analyzed the effect of BTI degradation on the performance of ring oscillators. In both device and circuit level, we observed that the effect of RDF on BTI degradation is more pronounced at a relatively lower BTI degradation. Since there are not enough experimental evidence that proves any impact of RDF on TDDB and SILC, further investigation in this regard is necessary. We conclude that the effect of RDF on the $V_{th}$ degradation needs to be properly addressed in a scaled CMOS technology and this may become crucial in accurate guard band estimation of a circuit.

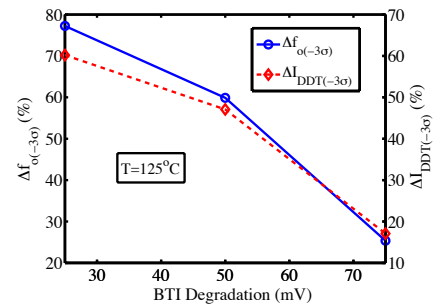

Fig. 15: -3$\sigma$ points of frequency of oscillation and dynamic supply current as a percentage of their corresponding standard deviations. Effect of RDF on BTI effect is more noticeable at lower BTI degradation.